\documentclass[conference]{IEEEtran}
\IEEEoverridecommandlockouts
\usepackage{cite}
\usepackage{amsmath,amssymb,amsfonts}
\usepackage{algorithm}
\usepackage{algorithmic}
\usepackage{multirow}
\usepackage{graphicx}
\usepackage{textcomp}
\usepackage{booktabs}
\usepackage{xcolor}
\usepackage{float} 
\def\BibTeX{{\rm B\kern-.05em{\sc i\kern-.025em b}\kern-.08em
    T\kern-.1667em\lower.7ex\hbox{E}\kern-.125emX}}
\begin{document}

\title{AV-GAN: Attention-Based Varifocal Generative Adversarial Network for Uneven Medical Image Translation\\
}

\author{\IEEEauthorblockN{Zexin Li}
\IEEEauthorblockA{\textit{SIGS of Tsinghua University} \\
Shenzhen, China \\
li-zx21@mails.tsinghua.edu.cn}
\and
\IEEEauthorblockN{Yiyang Lin}
\IEEEauthorblockA{\textit{SIGS of Tsinghua University} \\
Shenzhen, China \\
lyy20@mails.tsinghua.edu.cn}
\and
\IEEEauthorblockN{Zijie Fang}
\IEEEauthorblockA{\textit{SIGS of Tsinghua University} \\
Shenzhen, China \\
vison307@gmail.com}
\and
\IEEEauthorblockN{Shuyan Li}
\IEEEauthorblockA{\textit{Department of Automation of Tsinghua University} \\
Beijing, China \\
li-sy16@tsinghua.org.cn}
\and
\IEEEauthorblockN{Xiu Li}
\IEEEauthorblockA{\textit{SIGS of Tsinghua University} \\
Shenzhen, China \\
li.xiu@sz.tsinghua.edu.cn}
}

\maketitle

\begin{abstract}
Different types of staining highlight different structures in organs, thereby assisting in diagnosis. However, due to the impossibility of repeated staining,  we cannot obtain different types of stained slides of the same tissue area. Translating the slide that is easy to obtain~(e.g., H\&E) to slides of staining types difficult to obtain (e.g., MT, PAS) is a promising way to solve this problem. However, some regions are closely connected to other regions, and to maintain this connection, they often have complex structures and are difficult to translate, which may lead to wrong translations. In this paper, we propose the Attention-Based Varifocal Generative Adversarial Network (AV-GAN), which solves multiple problems in pathologic image translation tasks, such as uneven translation difficulty in different regions, mutual interference of multiple resolution information, and nuclear deformation. Specifically, we develop an Attention-Based Key Region Selection Module, which can attend to regions with higher translation difficulty. We then develop a Varifocal Module to translate these regions at multiple resolutions. Experimental results show that our proposed AV-GAN outperforms existing image translation methods with two virtual kidney tissue staining tasks and improves FID values by 15.9 and 4.16 respectively in the H\&E-MT and H\&E-PAS tasks.
\end{abstract}

\begin{IEEEkeywords}
GAN, Virtual Staining, Image Translation
\end{IEEEkeywords}

\section{Introduction}

In computational pathology, pathologists usually use pathological slides to obtain an accurate diagnosis and pathological staining is the gold standard\cite{montedonico2011histochemical}. Histochemical staining detects the presence of specific antigens in pathological tissues through the combination of antibodies and antigens and pathologists usually need to use different types of histochemical staining to achieve an accurate pathological diagnosis\cite{rivenson2020emerging}. For example, different types of pathological stains contain different molecular information, pathologists need several types of stainings to observe each specific structure of the kidney~\cite{liu2020predict,wang2020diagnostic}. For example, Masson trichromatic staining (MT) slides are for connective tissue observation, and Periodate Schiff Staining (PAS) slides are for the observation of the basement membrane. However, the tissue cannot be stained multiple times, as chemical staining destroys the cell structure\cite{roy2020collection}. To obtain a variety of staining types, existing methods apply virtual staining technology based on image translation\cite{de2021deep}. Specifically, they translate the staining slide which is easy to obtain (e.g. H\&E) to slides that require more preparation cost (e.g. MT, and PAS).

\begin{figure}[!t]
    \centering
    \includegraphics[width=0.7\linewidth]{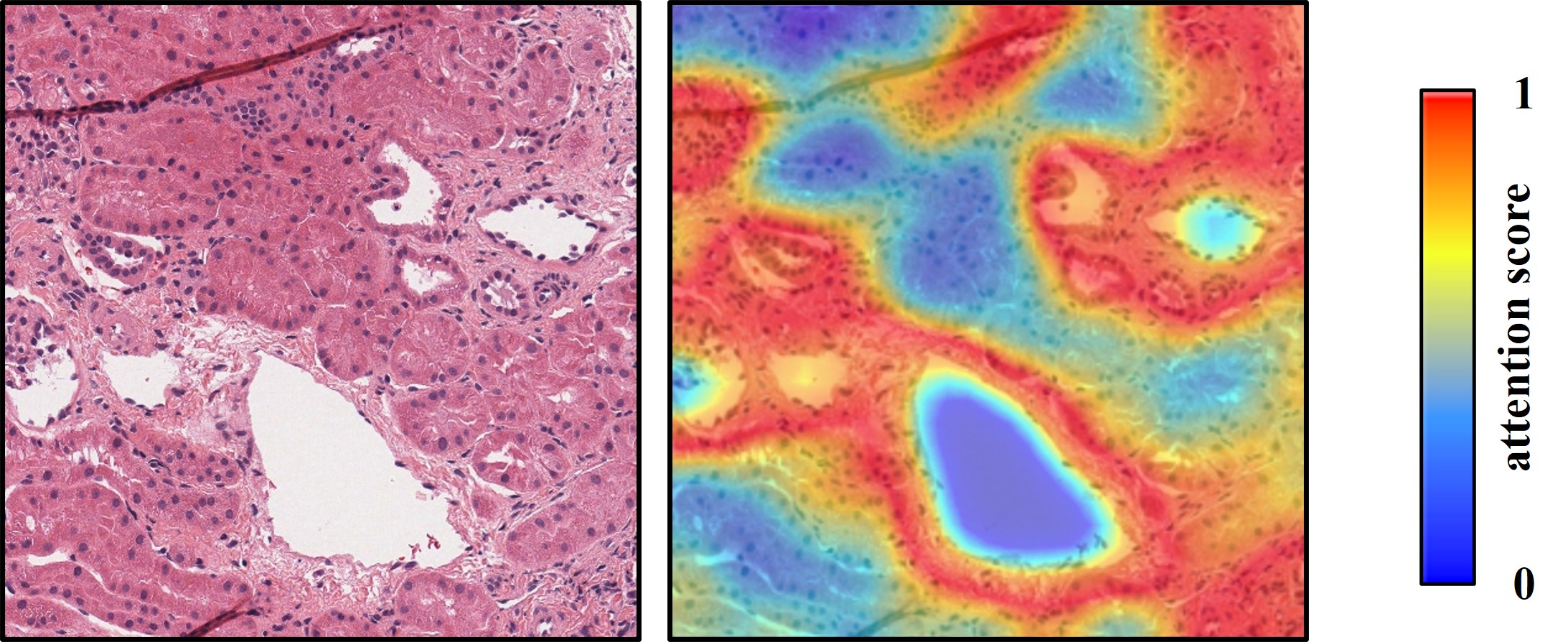}
    \caption{A H\&E stained patch and its attention map. The attention map shows that different regions deserve different extents of attention and structures like the edge of cavities in this patch deserve more attention, which corresponds to the fact that the edge is crucial for the shape of translated images.}
    \label{Figure2: The structure of Attention Module.}
\end{figure}

Although existing virtual staining technologies have achieved promising performance, there are still several challenges. Firstly, as shown in Figure 1, unlike the attention mechanism target in the previous method, which involves finding foreground objects to focus on in the background (such as finding horses in grasslands), in the task of pathological slides virtual staining, we need to find key regions that interact closely with tissues in other areas, as these key regions often require more complex structures to maintain this interaction, leading to increased translation difficulty\cite{galizia2001role}. For example, in the translation task of renal pathology slides, to efficiently filter the original urine sent from other tissues in a limited space, the glomerulus in the kidney must increase the contact area with the renal tubules, forming a complex cystic structure, making it difficult to translate. Motivated by this point, we design a novel attention mechanism that focuses on finding regions in pathological slides that are closely related to other regions and treating them as regions with higher translation difficulty.

After that, the existing methods have not been able to effectively translate information of different resolutions separately, making them unable to achieve the desired translation effect for information with high resolution (such as detailed texture, glomerular basement membrane, etc.) and information with low resolution (such as the shape of the tissue and the color appearance) simultaneously and separately\cite{kim2021fre}. Besides, the relative position of the nucleus may change during the translation of some models, which may lead to the deformation of some regions.

In this paper, we propose an Attention-Based Varifocal Generative Adversarial Network (AV-GAN) for image translation. To provide sufficient attention to the key regions (such as the glomerulus mentioned above) in the image, we adopt the attention mechanism to effectively focus on key regions. Besides, to better translate low-resolution features like the global shape as well as the appearance of the tissue and high-resolution features such as the detailed texture and local edge, we utilize a twin network architecture called varifocal module to achieve the extraction of features at different resolutions and the interaction between these features. In addition, we set the H channel loss to constrain the position of the nucleus to ensure that the shape of the tissue does not change significantly. Our work may do good for several downstream tasks such as object detection\cite{chu2023adversarial}. 

To sum up, our main contributions are as follows:
\begin{itemize}
     \item We design an Attention-based Key Region Selection Module based on the attention mechanism, which can select the structures in the tissue with higher translation difficulty and more connections to other areas.
    \item We propose a Varifocal Module, utilizing two generators to deal with global (low-resolution) and local (high-resolution) information respectively to ensure that information with different resolutions will not interfere with each other.
    \item We conduct experiments on the image translation tasks H\&E-MT and H\&E-PAS. Experimental results show that our method can translate H\&E staining slides into high-quality MT staining slides and PAS staining slides.
\end{itemize}

\section{Related Work}

\subsection{Virtual Staining in Histopathological Analysis}

Researchers have begun to explore virtual staining through image translation models. Lin et al. use StarGAN\cite{choi2020stargan} as the baseline to convert H\&E slides into special staining slides to achieve one-to-many image translation\cite{lin2022unpaired}. Yang et al. convert the H\&E staining slides of the kidney into special staining slides\cite{yang2022virtual}. Rivenson et al. virtually generate PAS, MT slides from H\&E slides. However, this model needs to input pixel-level registered data. Due to the large size of pathological WSIs, it is difficult to implement pixel-level registration\cite{rivenson2019virtual}. For the existing computational pathology image translation model, due to the differences in the translation difficulty of different regions of the slide (for example, the glomerulus is the key structure in kidney slides that should be focused on more), there is still a lack of an effective attention mechanism to enable the model to focus on these regions. Meanwhile, the input of high-resolution images forces the model to pay attention to details (such as tissue edges and cavities), while the input of low-resolution images forces the model to observe the global information and then handle the spatial relevance (such as the law of nuclear distribution). The existing pathology slides translation model can hardly explore the relationship between multiple resolutions.


\subsection{Generative Adversarial Network}

The Generative Adversarial network\cite{isola2017image} has become an important framework in image translation\cite{park2020contrastive}\cite{liu2019few}\cite{choi2020stargan}. Zhu et al. propose Pix2Pix\cite{isola2017image} and CycleGAN\cite{zhu2017unpaired} for supervised and unsupervised image translation. Kim et al. propose UGATIT\cite{kim2019u} for style transfer. Huang et al. propose MUNIT\cite{huang2018multimodal} to decouple image style information and content information. Mohammed et al. propose an image translation framework called AI-FFPE\cite{bengisu2021deep}. Zhang et al. design the Self-Attention Generative Adversarial Network (SAGAN)\cite{zhang2019self}, utilizing prompts from all feature locations to generate details. However, this scheme fails to pay attention to image information at different resolutions. Apart from that, Hu et al. propose the QS-Attn mechanism\cite{hu2022qs}. Among the features encoded by Encoder, the features that the author needs to constrain are filtered by the ``query-select'' method. However, this method only filters at the feature level and does not directly notice the regions in the image that should be concerned about. Zhang et al. successfully introduce the attention mechanism into GAN\cite{zhang2021dense}, but it is not applied in the image translation task. Kwon et al. propose a new layered adaptive diagonal spatial attention (DAT) layer\cite{Kwon_2021_ICCV}, which hierarchically processes spatial content and style, but this scheme does not process high-resolution and low-resolution information separately. In a word, the above frameworks fail to pay attention to the multi-resolution information, resulting in some key structures (such as the edge of tissue) not being well translated.

\section{Method}

In medical images, to solve the problem some structures are relatively complex (such as glomerulus, renal tubule, etc.) and cannot be translated precisely using a simple network, as shown in Figure 2, we propose an Attention-Based Varifocal Generative Adversarial Network (AV-GAN), which solves the problem of uneven difficulty distribution in image translation.

\begin{figure*}[!h]
    \centering
    \includegraphics[width=0.9\textwidth]{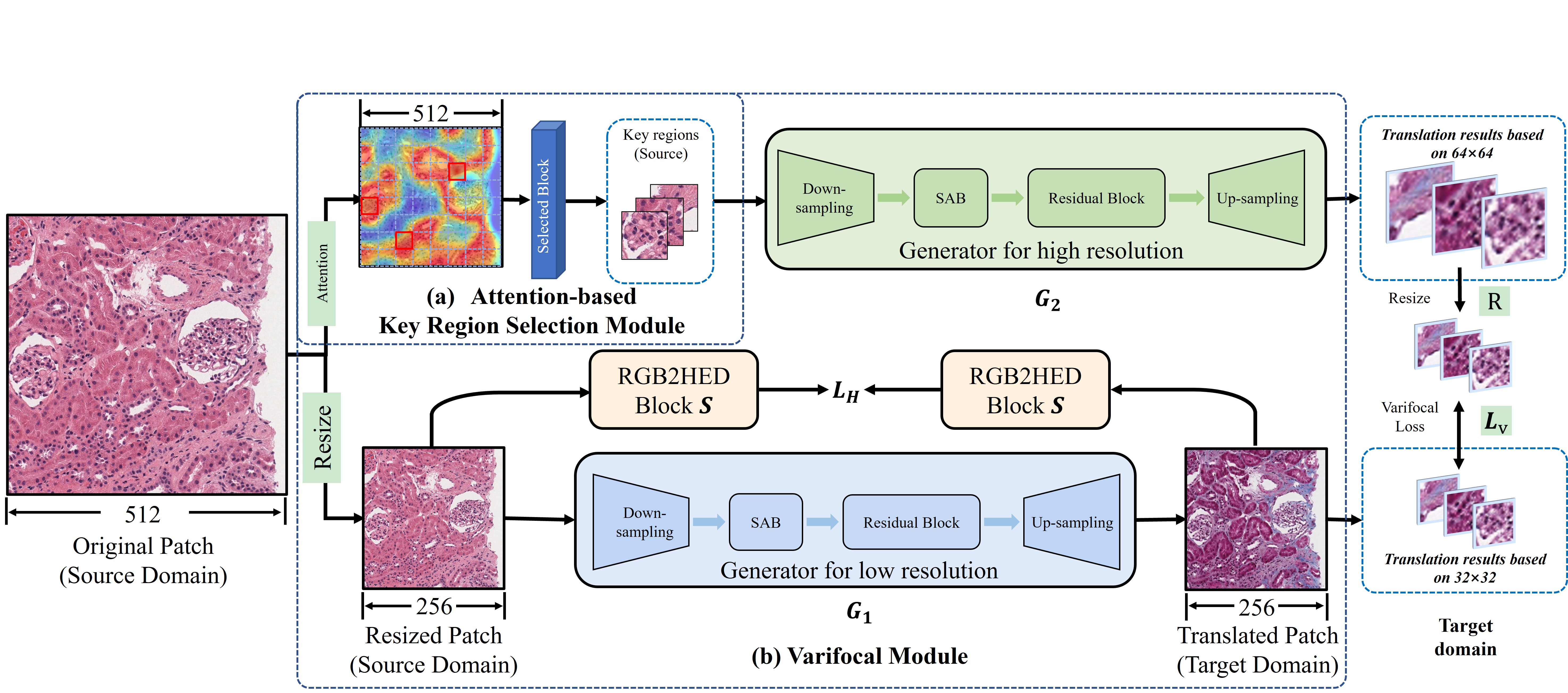}
    \caption{The structure of AV-GAN. $G_1$ and $G_2$ refer to the low-resolution generator and the high-resolution one respectively. $D_1$ and $D_2$ are the discriminators of high-resolution and low-resolution, which are not drawn in the figure. The Attention-Based Key Region Selection Module selects the region that is worth attention and the RGB2HED Block converts the RGB image to an HED image, whose H channel (nuclear channel) can be constrained. }
    \label{Figure1: The structure of AV-GAN.}
\end{figure*}

\begin{figure}[h]
    \centering
    \includegraphics[width=1.0\linewidth]{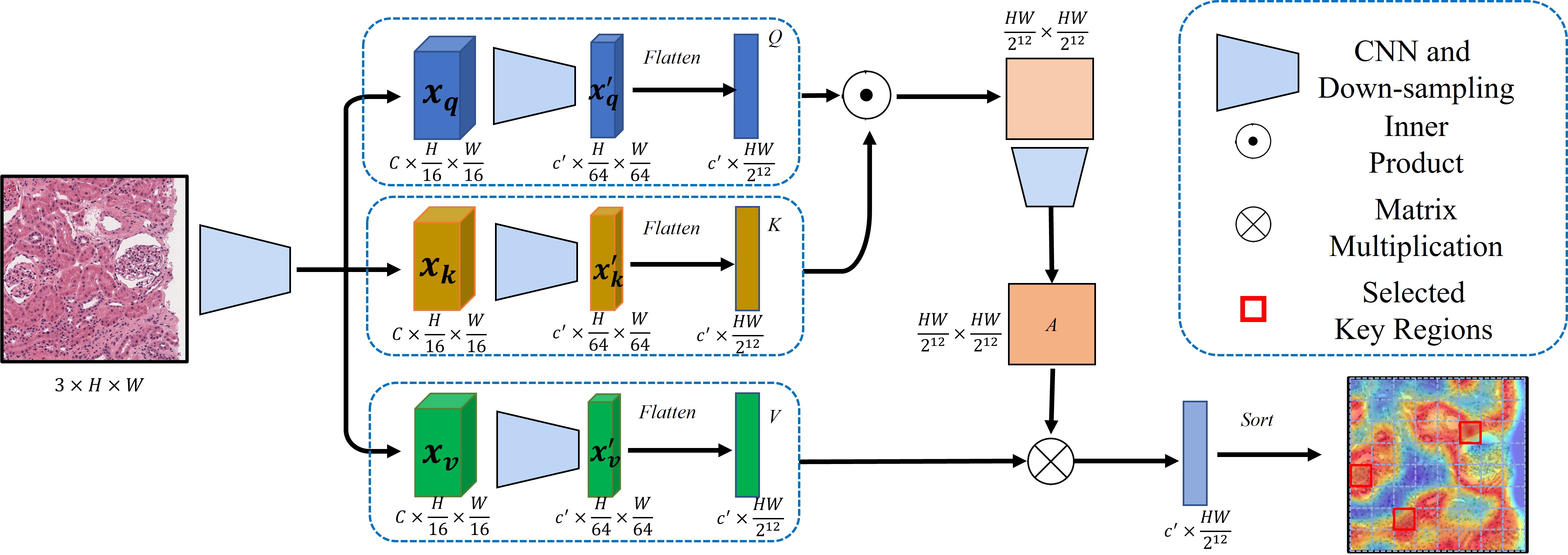}
    \caption{The structure of the attention module. The $Q$, $K$, and $V$ refer to the Query, Key, and Value network. The matrix $A$ represents the attention map, which is calculated through $Q$ and $K$. The network will select the region with key structures that need to be translated more precisely and return a cord list with each coordinate group representing a region.}
    \label{Figure2: The structure of Attention Module.}
\end{figure}

\subsection{Attention-based Key Region Selection Module}
Correct virtual staining of key structures is of great significance for accurate diagnosis. For example, in PAS slides, the glomerular basement membrane is stained evidently, so the glomerulus is the key structure of the slide. Correct staining of the glomerulus helps the diagnosis and treatment of a variety of kidney diseases (such as the classification of nephritis). To capture the key structures in pathological slides, we develop an Attention-based Key Region Selection Module, as shown in Figure 3. Firstly, we feed the input image $x \in \mathbb{R}^{C \times H \times W}$ into three independent convolution neural networks and downsampling blocks, getting the output $x_q$, $x_k$, and $x_v$ $\in \mathbb{R}^{C\times \frac{H}{16} \times \frac{W}{16}}$ where $x_q$, $x_k$, and $x_v$ are the representations of the image in the Query, Key and Value space, respectively, when $C$, $H$ and $W$ refer to the number of channels, the height, and width of a patch. To avoid the excessive size of the attention map, we downsample these representations to $x_q^\prime$, $x_k^\prime$, and $x_v^\prime$. Then, to obtain the strength of interaction between different regions, the three representations are flattened and the attention map of the image is calculated by 
\begin{equation}
  A = softmax({x_q^\prime}^{T}{x_k^\prime}),  
\end{equation}
where the softmax function is applied to the second dimension. 

Next, a matrix $P$, the importance of each key region for selection, is calculated by 
\begin{equation}
P = A{x_v^{\prime T}}.
\end{equation}
We take the average of $C$ channels for P and continue the following operation. At last, we sort the values in $P$ to find the top $n$ maximum values where $n$ is the hyperparameter that represents the number of key regions. Finally, we obtain the key regions $a_1$, $a_2$, ..., $a_i$, ..., $a_n$ ($1\leq i \leq n$) based on $P^\prime$ generated by $P$, and the generated process is described in the next paragraph. 

To transmit gradients to the attention module, we propose a novel attention mechanism, which is a key novelty of our work. After getting the attention map, the selected block calculates $m=sigmoid(1000(P-\theta))$, where $P$ is the average attention value of each region and $\theta$ determines how many regions to focus on (i.e.,  $n$). The larger the value of the parameter $\theta$, the fewer regions of interest can be selected and the less $n$ is. Then, we calculate the $P^\prime = m \times P$, and finally, the selected block selects the key regions for the varifocal module. The network also adds Spatial Attention Block which has an impact like \cite{bengisu2021deep}.

\subsection{Varifocal Module}

To improve the translation effect of the key regions selected in section III.A, we further propose a varifocal module. The module completes image translation at different resolutions through two independent generators, with each of them interacting with the other through varifocal loss, thus taking advantage of image features at different resolutions. 

The low-resolution generator ${G_1}$ consists of a down-sampling module, a spatial attention module, several bottleneck residual blocks, and an upsampling block. The main function of the low-resolution generator is to focus on the global characteristics of the image such as the shape and color style of the tissue. The low-resolution generator achieves better performance for translating the overall appearance of the image, especially the color styles, into another type of staining. 

At the same time, we set up a high-resolution generator ${G_2}$, which can better focus on local characteristics such as the local texture and the glomerulus. The high-resolution generator achieves better performance for translating the local structures of the image, especially the local contour of tissues and detailed texture, into another type of staining. 

To sum up, the low-resolution generator is better at capturing the global features such as the whole appearance, while the high-resolution generator is better at extracting local features such as some detailed textures. These two generators perform image translation at different resolutions and play a crucial role in the process of stain translation. we impel the generators of different resolutions to interact with each other properly instead of disorderly interference. The module fuses the advantages of two generators using L1 loss. The interaction process can be described as: 

\begin{equation}
\sum\limits_{n = 1}^N {{{\left\| {{C_n}({G_1}(I)) - {G_2}({a_n})} \right\|}_1}}
\end{equation}
In the formula, $I$ is the resized patch, $a_n$ is the ${n^{th}}$ key region, ${G_1}( \cdot )$ is the low-resolution generator, ${G_2}( \cdot )$ is the high-resolution generator, and $N$ is the total number of key regions. ${C_n}( \cdot )$ is the operation that crops the region from 
the resized patch $I$ corresponding to ${{a_n}}$, ${\left\|  \cdot  \right\|_1}$ is L1 loss. 

As a result, the fused network can deal with features in both high and low resolutions, capturing the information both globally and locally. Then we apply Varifocal loss to them as shown in Figure 2. In short, our model can significantly improve the virtual staining performance of key regions.

\subsection{Discriminator}


We use two discriminators $D_1$ and $D_2$ to distinguish low-resolution images and high-resolution local areas respectively. According to the research results of \cite{yang2022improving}, the structure of the generator and discriminator equipped in networks should match the difficulty of the task. Therefore, we choose the PatchGAN discriminator. 

\subsection{RGB2HED}

HED is a color space similar to RGB and is commonly used in computational pathology. RGB2HED is the process of transforming an image from RGB space to HED space by a stain deconvolution matrix. In HED space, the H channel highlights the nucleus. If the consistency of the H channel before and after translation can be maintained, the position of the nucleus can be guaranteed, and the structural precision after image translation is guaranteed. If we remove this constraint, we obtain FID values of 80.66 (H\&E-MT translation) and 97.85 (H\&E-PAS translation), worse than before, demonstrating the effectiveness of the H channel. Please refer to \textbf{H channel loss}.

\subsection{Loss function}

Our loss function has five terms and they are Adversarial loss $L_{adv}$, Identity loss $L_{idt}$, PatchNCE loss $L_{NCE}$, H channel loss $L_H$, and Varifocal loss $L_{v}$. Here we only introduce $L_H$ and $L_{v}$. You may refer to other terms in \cite{zhu2017unpaired,bengisu2021deep,park2020contrastive}.

\subsubsection{H channel loss}
To further constrain the spatial position relationship of the nucleus and ensure that the cell position distribution in the patch does not receive too much influence, we utilize the H channel separation operator $S$ to convert the source domain and generated target domain images into the HED color space and constrain the H channel of the HED color space. $G$ refers to the generator and $\mathbf{X}$ refers to the source domain. The H channel loss can be calculated by

\begin{equation}
L_H = ||S(\mathbf{X}) - S(G(\mathbf{X}))||_1.
\end{equation}

\subsubsection{Varifocal loss}
To ensure the consistency of the generated images between different resolutions, we set the Varifocal loss $L_v$ to constrain the difference in the generated results between the low-resolution and high-resolution generators. We find the corresponding regions of the selected regions in the low-resolution generated image and constrained those regions with the generation results of the high-resolution generator. To constrain images translated by the high-resolution generator and the low-resolution generator on the same scale, the larger image of the two is resized to the smaller image with the resizing operator $R$. The Varifocal loss is calculated by 
\begin{equation}
{L_v} = \sum_{i=1}^{n} || R(a_i) - b_i ||_1,
\end{equation}
where $a_i$ refers to the region mentioned in section III.A, and $b_i$ refers to the corresponding location in the translated image $G_1(x)$. 

\subsubsection{Optimization objective}
The optimization objective function can be calculated by

\begin{align}
L = & \lambda_{adv}L_{adv} + \lambda_{idt}L_{idt} + \nonumber \\ & \lambda_{NCE}(L_{NCE}(\mathbf{X}) + L_{NCE}(\mathbf{Y})) + \lambda_{H}L_{H} + \lambda_{v}L_{v},
\end{align}

where the weights $\lambda_{adv}$, $\lambda_{idt}$, $\lambda_{NCE}$, $\lambda_{H}$, and $\lambda_{v}$ are set to 1.0, 0.03, 1.0, 2.5, and 1.0 respectively to keep the loss terms in similar magnitude.

\section{Experiments and Results}

\subsection{Dataset and Data Preprocessing}
We have evaluated our proposed model over the H\&E-MT and H\&E-PAS image translation datasets of the ANHIR\cite{9058666}. In the ANHIR dataset, there are 5 groups of high-resolution tissue slides of human kidneys. Each of which has 3 unpaired Whole Slide Images (WSIs) stained with different kinds of stain, namely Hematoxylin-eosin (H\&E), Masson’s trichrome(MT), and Periodic Acid-Schiff (PAS). We divide the dataset according to the existing work\cite{lin2022unpaired} and segment all slides into 512 $\times$ 512 size patches. We adopt the overlapping measure in the segmentation process, which sets the stride of segmentation to 64. In the end, we obtained 17,964 H\&E patches, 20,139 MT patches, and 16,048 PAS patches as the training set 3,013 H\&E patches, 3,577 MT patches, and 2,997 PAS patches as the testing set.






\subsection{Experimental Details and Evaluation Criteria}
During training, we use Adam\cite{kingma2014adam} optimizer with $\beta_1=0.5$, $\beta_2=0.999$. The initial learning rate is set to 0.0002. The batch size is set to 1. We adopt Fr$\acute{e}$chet Inception Distance (FID)\cite{DBLP:journals/corr/HeuselRUNKH17}, Kernel Inception Distance (KID)\cite{Binkowski2018DemystifyingMG}, and Collaborative Structural Similarity (CSS)\cite{lin2022unpaired} to evaluate our model. The smaller the FID and KID values, the more similar the translated images are to real images. However, CSS only measures the similarity between the translated image and the source image, and higher CSS does not mean higher quality of the generated image. 



\subsection{Quantitative Results}
\subsubsection{Comparative Experiments of H\&E-MT Task}
We test 3013 H\&E images in the dataset and compare the results with starGAN\cite{choi2020stargan}, MW-GAN\cite{cao2019multi}, UGATIT\cite{kim2019u}, CycleGAN\cite{zhu2017unpaired}, AI-FFPE\cite{bengisu2021deep}, SSPRVS\cite{zeng2022semi}, and UMDST\cite{lin2022unpaired}. The results are shown in Table \ref{table 1}.  
\begin{table}[h]
    \centering
    \caption{Comparison of the effect of AV-GAN image translation with other image translation methods (H\&E-MT image translation task). The values in bold are the best ones, and the results underlined are the second-best ones.}
    \label{table 1}
    \begin{tabular}{lrrr}
        \toprule
        Model  & FID$\downarrow$ & CSS$\uparrow$ & 100KID$\downarrow$ \\
        \midrule
        StarGAN(2017)   & 284.37 & 0.62 & 29.11   \\
        MW-GAN(2019)   & 194.32 & 0.64 & 18.18   \\
        UGATIT(2019)     &142.50&0.33&7.33        \\
        CycleGAN(2017)   &95.22&0.69&4.34         \\
        AI-FFPE(2022)    &\underline{89.58}&\textbf{0.81}&\underline{3.56}        \\
        UMDST(2022) & 94.43 & 0.69 & 4.27\\
        SSPRVS(2022)\cite{zeng2022semi} & 140.84 & 0.68 & 6.69\\
        AV-GAN  (ours) &\textbf{73.68}&\underline{0.75}&\textbf{2.41}        \\
        \bottomrule
    \end{tabular}

\end{table}

In Table 1, the generated images of our model have better quality, indicating that they are more similar to the target domain image. From the effect of the H\&E-MT image translation task, we can conclude that our model has further decreased the FID value compared with existing methods. The 100KID also decreases by 1.15 compared with the previous SOTA value. The larger the CSS value, the better the target domain image can maintain the object structure in the source domain. Our CSS value exceeds most image translation methods largely, which shows that our model can effectively maintain the structural information in the source domain. This ensures the authenticity of cell structure in image translation tasks.

\subsubsection{Comparative Experiments of H\&E-PAS Task}

\begin{table}[h]
    \centering
    \caption{Comparison of the effect of AV-GAN image translation with other image translation methods (H\&E-PAS image translation task). The values in bold are the best ones, and the results underlined are the second-best ones.}
    \label{table 2}
    \begin{tabular}{lrrc}
        \toprule
        Model  & FID$\downarrow$ & CSS$\uparrow$ & 100KID$\downarrow$ \\
        \midrule
        StarGAN(2017)   & 219.71 & 0.60 & 20.48    \\
        MW-GAN(2019)   & 176.77 & 0.66 & 15.35   \\
        UGATIT(2019)     &156.88&0.31&10.11        \\
        CycleGAN(2017)   &106.51&0.66&3.95         \\
        AI-FFPE(2022)    &97.03&\textbf{0.77}&\underline{3.52}        \\
        UMDST(2022) & \underline{92.52} & 0.69 & 4.28\\
        SSPRVS(2022)\cite{zeng2022semi} & 131.46 & 0.68 & 7.21        \\
        AV-GAN (ours)&\textbf{88.36}&\underline{0.75}\textbf&\textbf{3.16}       \\
        \bottomrule
    \end{tabular}

\end{table}

Table \ref{table 2} shows the result of the H\&E-PAS image translation dataset, compared with other models, our method is significantly ahead. The FID value of the image dataset generated by our method is 88.36, which is better than other methods. One of the key reasons is that the basement membrane in H\&E patches is a detailed structure, and the traditional virtual staining image translation model can not recognize the structure well, resulting in a higher FID value. Our model also surpasses most comparison methods in terms of the CSS value, indicating that the structure information in the source domain is also well preserved. 

\subsection{Ablation Study}

\subsubsection{Sharing the parameters of the two generators $G_1$ and $G_2$}

We design the high-resolution generator $G_2$ to let the model focus more on details and avoid artifacts such as cell deformation in the translated images. We utilize the low-resolution generator $G_1$ to let the model get more global consideration and to avoid errors such as cell mass distribution displacement in the translated images. If the two generators share the parameters, the model can only focus on the common features rather than the features unique to each resolution. The result is shown in Table \ref{table 3}. The experiments show that using only one generator has obvious disadvantages over using two.


\begin{table}[h]
    \centering
    \caption{Compare the results of whether the weights of two generators are shared.}
    \label{table 3}
    \begin{tabular}{lcc}
        \toprule
        Measurement  & Shared parameters & Unshared parameters \\
        \midrule
        H\&E-MT FID$\downarrow$ & 128.07 & \textbf{73.68} \\
        H\&E-MT 100KID$\downarrow$ & 4.85 & \textbf{2.41} \\
        H\&E-PAS FID$\downarrow$ & 132.44 &  \textbf{88.36} \\
        H\&E-PAS 100KID$\downarrow$ & 5.04 &  \textbf{3.16} \\
        \bottomrule
    \end{tabular}

\end{table}

\subsubsection{Selecting the Proper Number of Key Regions}
To verify the importance of the Key Region Selection Module in region selection, we remove the Key Region Selection Module by selecting a fixed region, and then comparing the FID value and KID value of the generated image with the previous results. The results in Table \ref{table 4} show that when the number of regions is the same if the Key Region Selection Module is absent, the FID value and KID value will both increase, the quality of the generated image will decrease, and the distribution of the generated image dataset will be far from the distribution of the real target domain image dataset.


\begin{table}[h]
    \centering
    \caption{Ablation experiment. The quality of translated pictures in the case of ``using the attention mechanism to select region'' and ``selecting fixed region''. (H\&E-MT image translation task)}
    \label{table 4}
    \begin{tabular}{lrrc}
        \toprule
        Model  & FID$\downarrow$ & CSS$\uparrow$ & 100KID$\downarrow$ \\
        \midrule
        AV-GAN (1 attention region) &\underline{74.25}&\textbf{0.76}&\textbf{2.30}        \\
        AV-GAN (2 attention regions) &\textbf{73.68}&\underline{0.75}&\underline{2.41}        \\
        AV-GAN (3 attention regions) &76.01&0.73&2.51        \\
        AV-GAN (1 fixed region) &91.16&0.75&4.16        \\
        AV-GAN (2 fixed regions) &97.51&0.74&5.31        \\
        AV-GAN (3 fixed regions) &83.50&0.74&3.09        \\
        \bottomrule
    \end{tabular}

\end{table}

The table also shows the influence of the number of regions on evaluations. When the number of attention regions increases, the CSS value remains almost unchanged, and the FID and KID have decreased to varying degrees. The best FID value is 73.68, and the best KID value is 2.30, which is obtained when adding two regions and three regions, respectively. This shows that in the H\&E-MT image translation task, properly increasing the number of the attention regions is conducive to generating higher-quality images.

\begin{table}[h]
    \centering
    \caption{Ablation experiment. The quality of translated pictures in the case of ``using the attention mechanism to select region'' and ``selecting fixed region''. (H\&E-PAS image translation task)}
    \label{table 5}
    \begin{tabular}{lrrc}
        \toprule
        Model  & FID$\downarrow$ & CSS$\uparrow$ & 100KID$\downarrow$ \\
        \midrule
        AV-GAN (1 attention region) &\textbf{88.36}&\textbf{0.75}\textbf&\textbf{3.16}\\
        AV-GAN (2 attention regions) &90.20&0.73&3.83        \\
        AV-GAN (3 attention regions) &91.89&0.74&3.50        \\
        AV-GAN (1 fixed region) &95.68&0.75&3.41        \\
        AV-GAN (2 fixed regions) &96.19&0.73&3.43        \\
        AV-GAN (3 fixed regions) &97.96&0.73&3.99        \\
        \bottomrule
    \end{tabular}

\end{table}

The $n$ needs not to be too large (generally less than 3). In a picture, only key areas such as dense nuclei need to be selected. Once the key areas are selected, other key areas of the same type need not be selected again, because they have similar characteristics due to the homogeneity of pathological images. If an excessively large $n$ is chosen, the effect will not improve significantly except for increasing training time and GPU memory.

Table \ref{table 5} shows the results of the H\&E-PAS image translation task, which further validates the conclusions drawn previously. Specially, when the number of attention regions is 1, all evaluations are optimal. The result shows that because the overall color style of PAS is close to that of H\&E (both of them are soft pink), there is no need for too much attention. As a result, we can translate the H\&E image into the PAS image without adding too many attention regions.

\subsubsection{Selecting a Proper Size for Key Regions}

\begin{table}[h]
    \centering
    \caption{Comparative experiment. The quality of translated pictures in the case of ``selecting the region of 128×128'' and ``selecting the region of 64×64''. (H\&E-MT image translation task)}
    \label{table 6}
    \begin{tabular}{lrrc}
        \toprule
        The region size  & FID$\downarrow$ & CSS$\uparrow$ & 100KID$\downarrow$ \\
        \midrule
        64 (1 attention region) &\underline{74.25}&\textbf{0.76}&\textbf{2.30}        \\
        64 (2 attention regions) &\textbf{73.68}&\underline{0.75}&\underline{2.41}        \\
        64 (3 attention regions) &76.01&0.73&2.51      \\
        128 (1 attention region) &80.46&0.75&2.46        \\
        128 (2 attention regions) &89.49&0.73&3.52        \\
        128 (3 attention regions) &114.16&0.74&4.13        \\
        \bottomrule
    \end{tabular}

\end{table}

To further explore the selection strategy of the number of Key Regions and region size, we adjust their values on the H\&E-MT image translation task, which contains more structure and color conversion compared with the H\&E-PAS task. To be specific, we experiment with the region size of 64 and 128 and with the number of Key Regions of 1,2, and 3. The results are shown in Table \ref{table 6}. It can be found that when the region size is small, the performance will improve with the increase in the number of Key Regions. On the contrary, when the region size is large, the number of Key Regions should not be too large (Line 6). In summary, when the number of Key Regions and region size reach a balance, better performance can be achieved. Besides, the CSS values are almost all distributed around 0.75, which indicates that the number and size of the regions do not affect the CSS values.

\subsection{Evaluation of Experts}
To confirm the image translation effect of AV-GAN, we have designed a Visual Turing test performed by three experts. Since experts often see real stained slides, they have sufficient experience to determine whether a generated patch is realistic. On each task of H\&E-MT and H\&E-PAS image translation, each patch is translated with 7 models, generating 7 results and each expert is asked to select the most vivid one. We calculate the average of their results. Among the 4,000 best-translated MT patches, 1181.33 patches come from our model. Among the 4,000 best-translated PAS patches, 1149 patches come from our model. The results are shown in the Table \ref{table 7}.


\begin{table}[h]
    \centering
    \caption{The evaluation made by the experts. The numbers in the table indicate how many of the 4000 images with the best staining effect come from the corresponding method. The values in the table are the average values evaluated by three experts.}
    \label{table 7}
    \begin{tabular}{lrrc}
        \toprule
        Method  & MT patches & PAS patches  \\
        \midrule
        StarGAN (2017) & 289.67 & 236.33\\
        MW-GAN (2019) & 310.33 & 263.67\\
        UGATIT (2019) & 476.67 & 486.33 \\
        CycleGAN (2017) & 523.33 & 593.33 \\
        UMDST (2022) & 493.67 & 468.67 \\
        AI-FFPE (2022) & 725.00 & 802.67 \\
        AV-GAN & \textbf{1181.33} & \textbf{1149} \\
        \bottomrule
    \end{tabular}

\end{table}

\subsection{Qualitative Experiment}
\begin{figure}[h]
    \centering
    \includegraphics[width=0.8\linewidth]{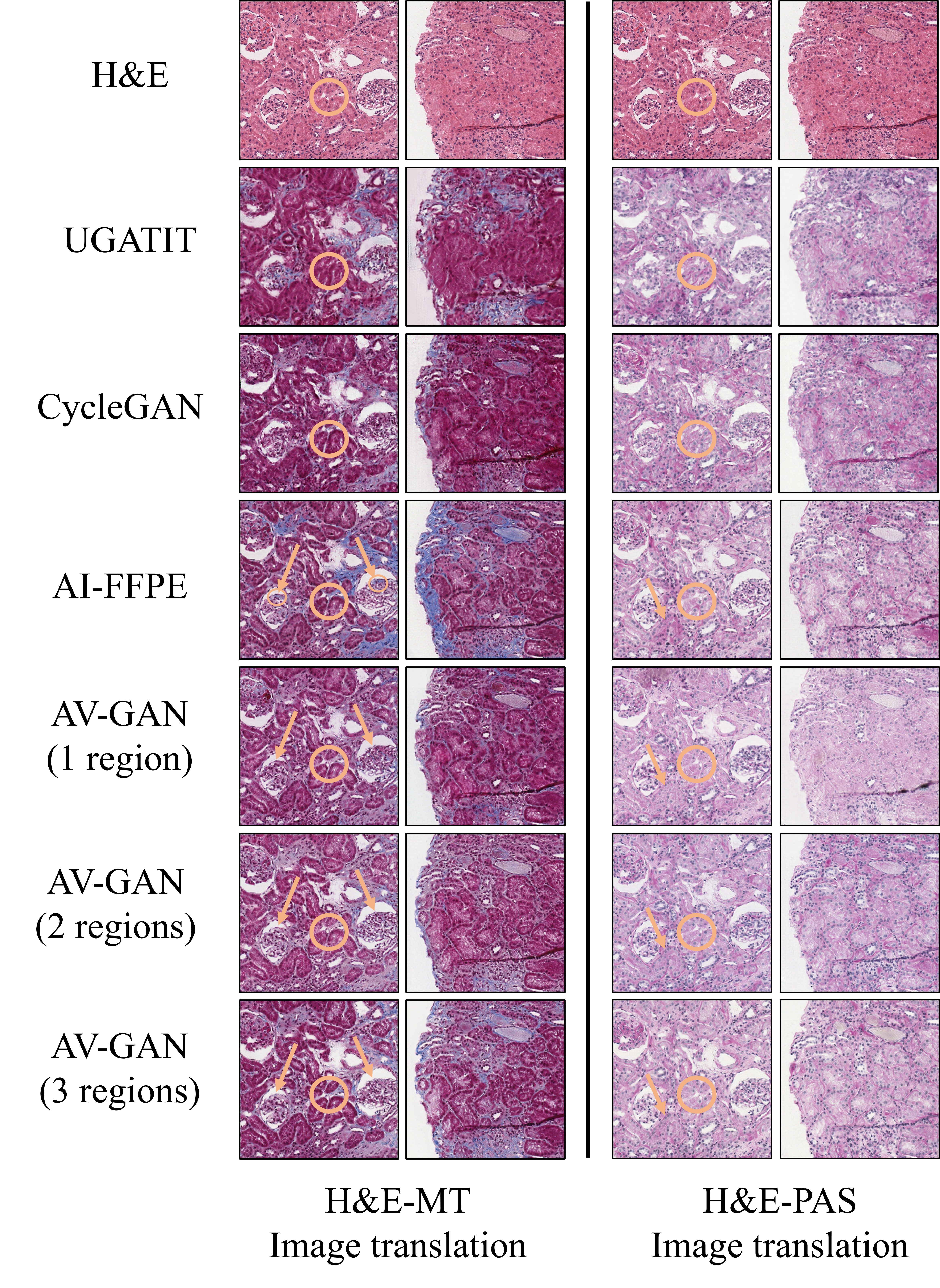}
    \caption{Image translation results. The first line is the original H\&E-stained images, the second line, the third line and the fourth line are the results of image translation using UGATIT, CycleGAN, and AI-FFPE models, and the fifth to seventh lines are the results of image translation using AV-GAN with different numbers of regions. Our task is to translate H\&E staining into MT staining and PAS staining.}
    \label{Figure3}
\end{figure}
As shown in the left side of Figure 4 and Figure 6, in the H\&E-MT image translation task, the result of  UGATIT is relatively poor, which not only has low legibility, but also has the phenomenon of nucleus loss, and the relatively close nuclei cannot be separated from each other (see the yellow circle in Figure 4). The translation of CycleGAN is slightly better, but the nucleus is hidden in the red zone and cannot be distinguished (see yellow circle in Figure 4). Then, the AI-FFPE model is slightly better in terms of structural clarity. Our AV-GAN can also ensure structural clarity and distinguish red muscle fibers better. Furthermore, in the part pointed out by the yellow arrow, there are two glomerular structures. According to the diagnosis of the pathologist, this patient did not suffer from the disease that could lead to the deposition of collagen fibers in the glomerulus and the MT staining slide of the glomerulus should not appear blue, but AI-FFPE translates some areas in the glomerulus into blue (Figure 5). 

\begin{figure}[h]
    \centering
    \includegraphics[width=0.8\linewidth]{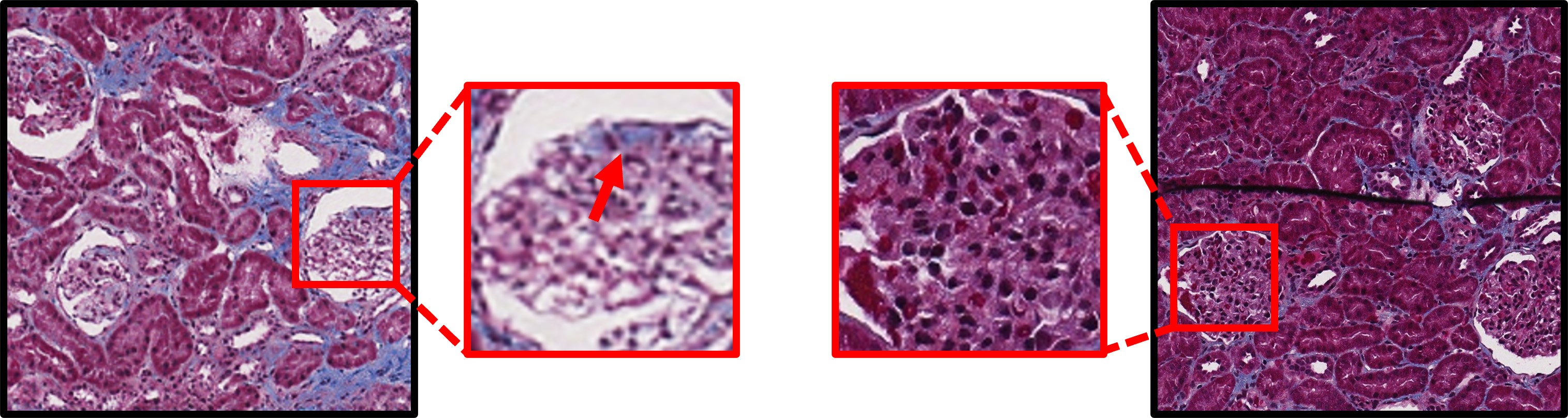}
    \caption{The leftmost and rightmost are MT patches stained by AI-FFPE and a real patch of the same patient's MT patches (the dataset is an unpaired dataset). A blue zone appears at the location the red arrow points at, while the right patch indicates that there should be no blue in the patient's glomerular.}
    \label{Figure4}
\end{figure}

\begin{figure*}[!h]
    \centering
    \includegraphics[width=0.80\textwidth]{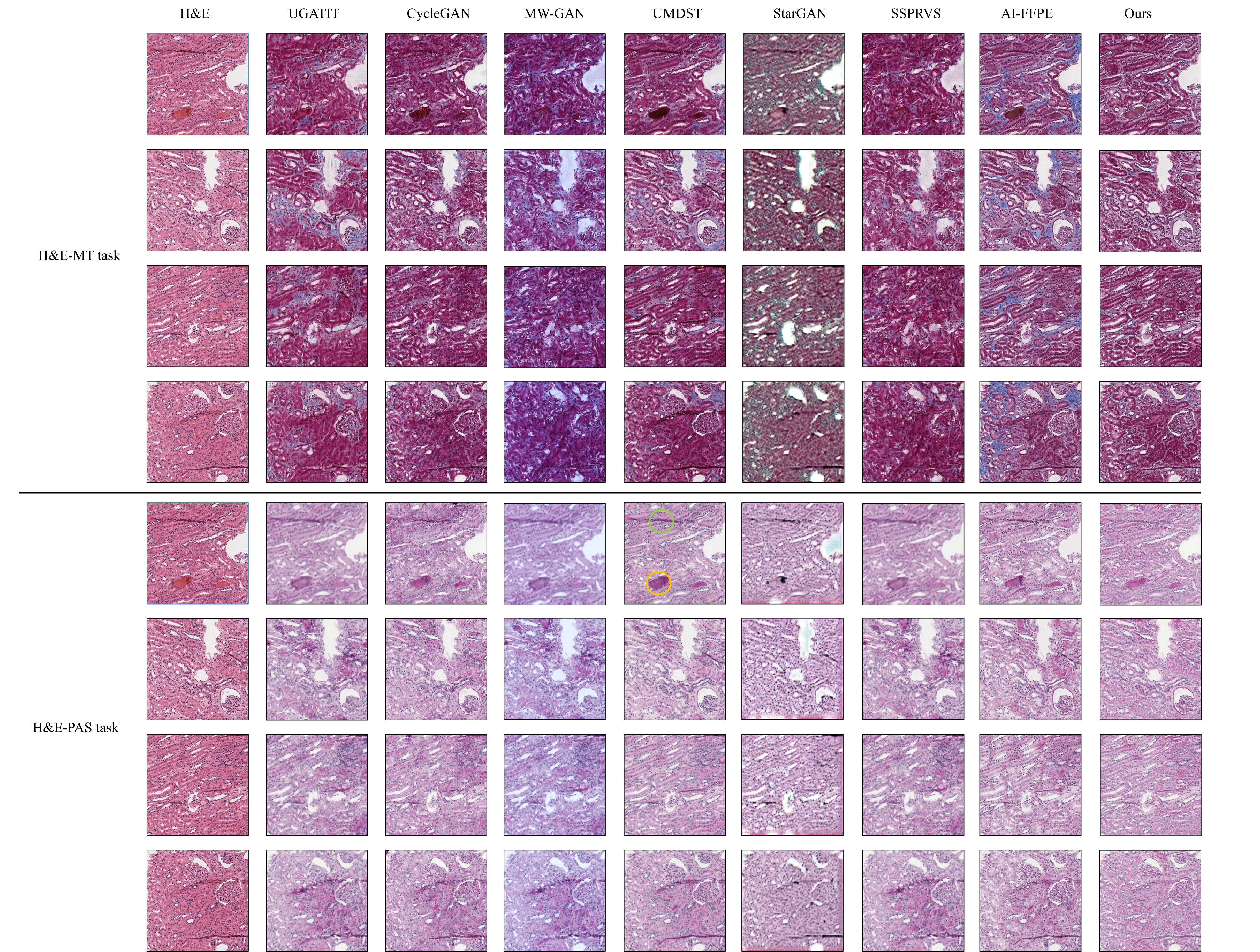}
    \caption{Results of more qualitative experiments.}
    \label{Figure5}
\end{figure*}

The right side of Figure 4 is the result of the image translation experiment of H\&E-PAS staining. Similar to the H\&E-MT staining image translation experiment, the image translation effect of UGATIT and CycleGAN is poor. The main function of PAS staining is to stain the glomerular basement membrane pink. In the part pointed out by the yellow arrow, AI-FFPE's result appears as a diffuse pink area, which is an error. After applying our model, the error is corrected.


\section{Conclusion}
We propose Attention-Based Varifocal Generative
Adversarial Network (AV-GAN), effectively solves the problems of uneven translation difficulty, interference of different resolution information, and nuclear deformation in pathological image translation. In AV-GAN, the Attention-based Key Region Selection Module allows the network to find the area that is most closely related to other regions, which is often the area with higher translation difficulty, and the Varifocal Module decouples the information of regions difficult to translate at multiple resolutions and translate them separately, which improves the translation quality of key regions. This paper also utilizes the H channel loss term to restricts the position distribution of the nucleus and ensures that the tissue morphology does not change after translation. Our model gets the state-of-the-art performance on H\&E-MAS and H\&E-PAS translation tasks. 

\bibliographystyle{unsrt}

\bibliography{sample-base}

\end{document}